\documentclass[preprint,showpacs,preprintnumbers,amsmath,amssymb]{revtex4}

\usepackage{graphicx}
\usepackage{dcolumn}
\usepackage{bm}

\DeclareMathOperator{\tr}{tr}
\DeclareMathOperator{\U}{U}
\DeclareMathOperator{\SU}{SU}
\DeclareMathOperator{\Real}{Re}
\newcommand{\rmd}{{\rm d}}

\begin{document}

\preprint{RIKEN-TH-107}

\title{Perturbative Spectrum of a Yukawa-Higgs Model with Ginsparg-Wilson
Fermions}%

\author{Hiroshi Suzuki}
\email{hsuzuki@riken.jp}
\affiliation{%
Theoretical Physics Laboratory, RIKEN, Wako 2-1, Saitama 351-0198, Japan
}%

\date{\today}

\begin{abstract}
A Yukawa-Higgs model with Ginsparg-Wilson (GW) fermions, proposed recently by
Bhattacharya, Martin and Poppitz as a possible lattice formulation of chiral
gauge theories, is studied. A simple argument shows that the gauge boson always
acquires mass by the St\"uckelberg (or, in a broad sense, Higgs) mechanism,
regardless of strength of interactions. The gauge symmetry is spontaneously
broken. When the gauge coupling constant is small, the physical spectrum of the
model consists of massless fermions, massive fermions and
\emph{massive\/} vector bosons.
\end{abstract}

\maketitle

Recently, Bhattacharya, Martin and Poppitz~\cite{Bhattacharya:2006dc} proposed
a Yukawa-Higgs model with GW fermions as a possible lattice formulation of
chiral gauge theories. (For reviews on various approaches on this problem, see
Refs.~\cite{Petcher:1993mn,Shamir:1995zx,Golterman:2000hr,Golterman:2004wd}.)
This approach was subsequently studied by analytical and numerical
methods~\cite{Giedt:2007qg,Poppitz:2007tu}. The idea~\cite{Bhattacharya:2006dc}
is that half the fermion sector  (``mirror fermions'') in a vector-like theory
decouples, forming heavy composite fermions by strong Yukawa interactions and,
at the same time, the gauge symmetry is not spontaneously broken by keeping the
Higgs sector in a symmetric phase (by choosing a coupling $\kappa$ small; see
below). They argued that, in this way, a desired pattern of spectrum as chiral
gauge theory, that is, massless Weyl fermions interacting via massless gauge
bosons, can be realized.

If this scenario comes true, it implies a great simplification because the
lattice chiral gauge theory formulated in
Refs.~\cite{Luscher:1998du,Luscher:1999un} on the basis of the GW relation
requires ingenious construction of the fermion integration measure. An
``ideal'' measure must be consistent with the locality, gauge invariance and
smoothness and its construction is far from being trivial. Although an explicit
way of construction is known for (anomaly-free) $\U(1)$ gauge
theories~\cite{Luscher:1998du,Kadoh:2003ii,Kadoh:2004uu,Kadoh:2005fa} (and for
the electroweak $\SU(2)_L\times\U(1)_Y$
theory~\cite{Kikukawa:2000kd,Kadoh:2007}), for general non-abelian theories the
way of construction has been known only to all orders of perturbation
theory~\cite{Luscher:2000zd}. (The existence of an ideal measure in
perturbation theory was shown in Refs.~\cite{Suzuki:2000ii,Igarashi:2000zi}.)
Construction of the fermion integration measure in a non-perturbative level is
a mathematically complex problem requiring, first of all, non-abelian
generalization of a local cohomology argument on the
lattice~\cite{Luscher:1998kn,Fujiwara:1999fi,Fujiwara:1999fj,%
Kikukawa:2001mw,Igarashi:2002zz,Kadoh:2003ii,Kadoh:2004uu,Kadoh:2005fa} that is
so far available only for the gauge group $\U(1)$. On the other hand, as we
will review below, the fermion integration measure in the proposal of
Ref.~\cite{Bhattacharya:2006dc} is quite simple. Therefore, there is hope such
that the mirror fermions and the Higgs field ``dynamically'' provide an ideal
integration measure of massless Weyl fermions while evading the above
complexity.

In this brief report, we show that the model unfortunately fails to meet above
expectations. The physical vector boson always acquires mass by the
St\"uckelberg (or Higgs) mechanism, regardless of strength of interactions. In
this sense, the gauge symmetry is always spontaneously broken. Our
argumentation to show this is very simple and kinematical. That is, it relies
only on a symmetrical structure of the model. Because of the simplicity of this
argument, we believe that some workers in this field have already arrived at
the conclusion identical to ours. In fact, it has been known that a compact
Higgs field (see below) can be interpreted as a St\"uckelberg field; see, for
example, Ref.~\cite{Golterman:1991re}. On the other hand, it appears that the
point we want to emphasize below is not so well-appreciated.

As an example, we take the so-called ``345'' model studied in
Ref.~\cite{Bhattacharya:2006dc}. The target theory is a two-dimensional $\U(1)$
chiral gauge theory that contains two left-handed Weyl fermions (their $\U(1)$
charges are 3 and~4, respectively) and one right-handed Weyl fermion (its
$\U(1)$ charge is 5). Since $3^2+4^2=5^2$, this system is free from the gauge
anomaly (the issue of the gauge anomaly plays no central role in what follows,
however). The partition function of the model, according to
Refs.~\cite{Bhattacharya:2006dc,Giedt:2007qg,Poppitz:2007tu}, is defined by
\begin{widetext}
\begin{equation}
   \mathcal{Z}=\int\prod_x\left(\prod_\mu\rmd U(x,\mu)\right)\,
   \rmd\phi(x)\,
   \left(\prod_{q=0,3,4,5}\rmd\psi_q(x)\,\rmd\overline\psi_q(x)\right)e^{-S},
\label{one}
\end{equation}
\end{widetext}
where $\mu$ runs from 0 to~1. In this expression, $U(x,\mu)$ denotes the
$\U(1)$ link variables and $\phi(x)\in\U(1)$ is a \emph{compact\/} Higgs field.
$\rmd U(x,\mu)$ and $\rmd\phi(x)$ are the Haar measures. There are four fermion
fields, $\psi_0(x)$, $\psi_3(x)$, $\psi_4(x)$ and~$\psi_5(x)$. The first
one~$\psi_0$ is a spectator having no $\U(1)$ charge and it is introduced to
form appropriate Yukawa interactions below. Note that the integration measure
of the fermions is trivial in a sense that it is a simple product of Grassmann
integrals (like that in lattice QCD). This point is quite different from
construction of the fermion integration measure in the framework of
Refs.~\cite{Luscher:1998du,Luscher:1999un} that requires a careful choice of
basis vectors in which the Weyl fermion fields are expanded. The total action
is given by
\begin{equation}
   S=S_{\text{G}}+S_\kappa+S_{\text{light}}+S_{\text{mirror}}.
\label{two}
\end{equation}
We do not need to specify an explicit form of the gauge action $S_{\text{G}}$,
although we assume that it belongs to a same universality class as the
plaquette action. What is important to us is its invariance under the lattice
gauge transformation ($\hat\mu$ denotes a unit vector in the $\mu$-direction
and the lattice spacing~$a$ is set to 1 in most part of this paper)
\begin{align}
   U(x,\mu)\to\Lambda(x)U(x,\mu)\Lambda(x+\hat\mu)^{-1},
\label{three}
\end{align}
where $\Lambda(x)\in\U(1)$. The kinetic term of the Higgs field $S_\kappa$ is 
\begin{equation}
   S_\kappa=\kappa\sum_x\sum_\mu\Real
   \left\{1-\phi(x)^{-1}U(x,\mu)\phi(x+\hat\mu)\right\},
\label{four}
\end{equation}
where we have assumed that the field~$\phi(x)$ has the $\U(1)$ charge $+1$.
The gauge transformation of $\phi$ is thus given by
\begin{align}
   \phi(x)\to\Lambda(x)\phi(x).
\label{five}
\end{align}
Of course, $S_\kappa$ is invariant under the gauge transformations (\ref{three})
and~(\ref{five}). The actions of ``light'' fermions, which correspond to
massless Weyl fermions in the target theory, are given by
\begin{widetext}
\begin{equation}
   S_{\text{light}}=\sum_x\left\{
   \overline\psi_{0,+}D_0\psi_{0,+}+\overline\psi_{3,-}D_3\psi_{3,-}
   +\overline\psi_{4,-}D_4\psi_{4,-}+\overline\psi_{5,+}D_5\psi_{5,+}
   \right\}
\label{six}
\end{equation}
and, for ``mirror'' ones
\begin{align}
   S_{\text{mirror}}&=\sum_x\left\{
   \overline\psi_{0,-}D_0\psi_{0,-}+\overline\psi_{3,+}D_3\psi_{3,+}
   +\overline\psi_{4,+}D_4\psi_{4,+}+\overline\psi_{5,-}D_5\psi_{5,-}
   \right\}
\nonumber\\
   &\quad{}+y\sum_x\bigl\{
   \overline\psi_{0,-}(\phi^{-1})^3\psi_{3,+}
   +\overline\psi_{3,+}(\phi)^3\psi_{0,-}
   +\overline\psi_{0,-}(\phi^{-1})^4\psi_{4,+}
   +\overline\psi_{4,+}(\phi)^4\psi_{0,-}
\nonumber\\
   &\qquad\qquad{}+
   \overline\psi_{3,+}(\phi^{-1})^2\psi_{5,-}
   +\overline\psi_{5,-}(\phi)^2\psi_{3,+}
   +\overline\psi_{4,+}(\phi^{-1})\psi_{5,-}
   +\overline\psi_{5,-}(\phi)\psi_{4,+}
   \bigr\}
\nonumber\\
   &\quad{}+h\sum_x\bigl\{
   \psi_{0,-}^TB(\phi^{-1})^3\psi_{3,+}
   -\overline\psi_{3,+}B(\phi)^3\overline\psi_{0,-}^T
   +\psi_{0,-}^TB(\phi^{-1})^4\psi_{4,+}
   -\overline\psi_{4,+}B(\phi)^4\overline\psi_{0,-}^T
\nonumber\\
   &\qquad\qquad{}+
   \psi_{3,+}^TB(\phi^{-1})^8\psi_{5,-}
   -\overline\psi_{5,-}B(\phi)^8\overline\psi_{3,+}^T
   +\psi_{4,+}^TB(\phi^{-1})^9\psi_{5,-}
   -\overline\psi_{5,-}B(\phi)^9\overline\psi_{4,+}^T
   \bigr\},
\label{seven}
\end{align}
\end{widetext}
where $B$ denotes the charge conjugation matrix in two dimensions.

The expressions~(\ref{six}) and~(\ref{seven}) need some explanation. The
subscript~$q$ of the lattice Dirac operators $D_q$ ($q=0$, 3, 4 or~5)
indicates the $\U(1)$ charge of the fermion it acts. In the lattice Dirac
operator~$D_q$, the link variables are contained with the representation
$(U(x,\mu))^q$. The Dirac operator $D_q$ must be gauge covariant. That is,
under the gauge transformation~(\ref{three}), it transforms as
$D_q\to(\Lambda)^qD_q(\Lambda^{-1})^q$. It is also assumed that $D_q$ satisfies
the GW relation~\cite{Ginsparg:1981bj}
\begin{equation}
   \gamma_5D_q+D_q\gamma_5=D_q\gamma_5D_q.
\label{eight}
\end{equation}
Neuberger's operator~\cite{Neuberger:1997fp,Neuberger:1998wv} is simplest among
such lattice Dirac operators. Defining the
combination~$\hat\gamma_{q,5}=\gamma_5(1-D_q)$, one has from the GW relation 
\begin{equation}
   (\hat\gamma_{q,5})^2=1,\qquad
   D_q\hat\gamma_{q,5}=-\gamma_5D_q
\label{nine}
\end{equation}
and hence $\hat\gamma_{5,q}$ is a lattice analogue of the
$\gamma_5$~\cite{Luscher:1998pq,Narayanan:1998uu,Niedermayer:1998bi}. We also
introduce projection operators
\begin{equation}
   \hat P_{q,\pm}={1\over2}(1\pm\hat\gamma_{q,5}),\qquad
   P_\pm={1\over2}(1\pm\gamma_5)
\label{ten}
\end{equation}
and define chiral components of lattice fermions by
\begin{equation}
   \psi_{q,\pm}(x)\equiv\hat P_{q,\pm}\psi_q(x),\qquad
   \overline\psi_{q,\pm}(x)\equiv\overline\psi_q(x)P_\mp
\label{eleven}
\end{equation}
for each~$q$. Note that, because of the property~(\ref{nine}), the action of a
lattice Dirac fermion completely decomposes into the right- and the left-handed
parts
\begin{equation}
   \overline\psi_q(x)D_q\psi_q(x)=
   \overline\psi_{q,+}(x)D_q\psi_{q,+}(x)+\overline\psi_{q,-}(x)D_q\psi_{q,-}(x).
\label{twelve}
\end{equation}
As emphasized in Refs.~\cite{Bhattacharya:2006dc,Giedt:2007qg,Poppitz:2007tu},
this complete chiral separation of a lattice action is peculiar to formulation
based on the lattice Dirac operator satisfying the GW relation. Since the Dirac
operator is gauge covariant, so are the projection operators,
$\hat P_{q,\pm}\to(\Lambda)^q\hat P_{q,\pm}(\Lambda^{-1})^q$
(and of course $P_\pm\to(\Lambda)^qP_\pm(\Lambda^{-1})^q$). Then the actions
(\ref{six}) and~(\ref{seven}) are clearly invariant under the simultaneous
gauge transformations (\ref{three}), (\ref{five}) and
\begin{equation}
   \psi_q(x)\to(\Lambda(x))^q\psi_q(x),\qquad
   \overline\psi_q(x)\to\overline\psi_q(x)(\Lambda(x)^{-1})^q.
\label{thirteen}
\end{equation}
The action for light fermions~$S_{\text{light}}$ is identical to the action of
the Weyl fermions that would be taken in the formulation of
Ref.~\cite{Luscher:1998du}. See also Ref.~\cite{Niedermayer:1998bi}.

The Yukawa interactions in Eq.~(\ref{seven}) are
chosen~\cite{Bhattacharya:2006dc} so that they break all global (vector as well
as chiral) $\U(1)$ transformations of mirror fermions,
$\psi_{0,-}$, $\psi_{3,+}$, $\psi_{4,+}$ and~$\psi_{5,-}$, except the global
$\U(1)$ part of the gauge transformations~(\ref{thirteen})
and~(\ref{five}).

Now, our argument is based on a simple change of integration variables in
Eq.~(\ref{one}). Instead of gauge variant original variables $U(x,\mu)$,
$\psi_q(x)$ and~$\overline\psi_q(x)$, one may use gauge \emph{invariant\/} ones
\begin{align}
   &U'(x,\mu)=\phi(x)^{-1}U(x,\mu)\phi(x+\hat\mu),
\nonumber\\
   &\psi_q'(x)=(\phi(x)^{-1})^q\psi_q(x),\qquad
   \overline\psi_q'(x)=\overline\psi_q(x)(\phi(x))^q.
\label{fourteen}
\end{align}
For any fixed configuration of~$\phi(x)$, the jacobian from
$\{U(x,\mu),\psi_q(x),\overline\psi_q(x)\}$ to
$\{U'(x,\mu),\psi_q'(x),\overline\psi_q'(x)\}$ is unity because
$\phi(x)\in U(1)$ and the numbers of integration variables $\psi_q(x)$ and
$\overline\psi_q(x)$ are same. It is obvious that the action~$S$, when
expressed in terms of these primed variables, does not contain the
$\phi$-field anymore. This is simply a reflection of the gauge invariance of
the action and the fact that the compact field $\phi(x)\in\U(1)$ can be
regarded as a parameter of the lattice gauge transformation. Then, since $\phi$
is compact, we can integrate it out from the partition function. 

After this change of variables, the kinetic term of the $\phi$-field becomes the
mass term of the (gauge invariant) vector boson~\footnote{The importance of
this phenomenon in a somewhat different context was stressed to me by Yoshio
Kikukawa.}
\begin{equation}
   S_\kappa=\kappa\sum_x\sum_\mu\Real\left\{1-U'(x,\mu)\right\}.
\end{equation}
Thus we see that the vector boson acquires mass by the St\"uckelberg (or, in
a broad sense, Higgs) mechanism~\footnote{Here, by the St\"uckelberg mechanism,
we mean the situation in which the gauge boson acquires mass by completely
absorbing all scalar fields and all the scalar fields correspond to gauge
degrees of freedom.}. (For a review on the St\"uckelberg mechanism, see
Ref.~\cite{Ruegg:2003ps}.) Our choice of the primed variables~(\ref{fourteen})
corresponds to the so-called unitary gauge and one can say that the gauge
symmetry is spontaneously broken. In terms of the primed variables, the Yukawa
interactions in $S_{\text{mirror}}$ become mass terms of mirror fermions. Note
that the above argument holds regardless of strength of interactions. In the
present two-dimensional theory, the dimensionless gauge coupling constant~$ag$
goes to zero in the continuum limit $a\to0$. For $ag\ll1$, the situation
relevant in the continuum limit, the spectrum of the model consists of massless
fermions, massive fermions and \emph{massive\/} vector bosons, interacting
through chiral couplings. The mass of the massive fermions is $O(y/a)$
or~$O(h/a)$. The mass of the vector boson is, on the other hand, $O(\kappa g)$.
Since the variables~(\ref{fourteen}) are gauge invariant, this is a physical
spectrum. This perturbative physical spectrum differs from the one, that might
be expected in chiral gauge theories in the perturbative regime.

In the above example, the Higgs field has the $\U(1)$-charge $+1$ and this
charge is, according to the terminology of Ref.~\cite{Fradkin:1978dv}, the
``fundamental representation''. In fact, our argument above is nothing but the
argument used in Ref.~\cite{Fradkin:1978dv} to show that lattice gauge models
with a compact Higgs field in the fundamental representation are in the Higgs
phase. The presence of fermions is not relevant in this argument. Here, one
cannot repeat an argument of Ref.~\cite{Fradkin:1978dv} which shows the
existence of the Coulomb phase (in which the gauge symmetry is not
spontaneously broken) for $\kappa\ll1$, because that argument is based on the
presence of a phase transition in pure gauge models. In two-dimensional gauge
models, such a phase transition does not occur.

A similar argument can be repeated for the two-dimensional ``1-0''
model~\cite{Giedt:2007qg,Poppitz:2007tu} that contains two fermions with the
$\U(1)$ charges $+1$ and $0$, respectively. The target chiral gauge theory of
this model is anomalous because $1^2\neq0$ but nevertheless our argument
proceeds without any essential change. We again have massless fermions, massive
fermions and massive vector bosons. This is very natural because
two-dimensional anomalous $\U(1)$ chiral gauge theory would be consistent, if
the vector boson is allowed to be massive~\cite{Jackiw:1984zi,Halliday:1985tg}.

In Refs.~\cite{Bhattacharya:2006dc,Giedt:2007qg,Poppitz:2007tu}, the authors
are considering the limit $ag=0$, where $ag$ is the dimensionless gauge
coupling constant, as a first approximation. Then they completely neglect the
gauge fields \emph{including\/} the gauge degrees of freedom. What we wanted
to emphasize in this note is that this kind of approximation which neglects the
underlying gauge symmetry can sometimes be misleading. In other words, the
nature of the spontaneous breaking of a continuous symmetry crucially depends
on whether the symmetry is global or local (i.e., gauged). For example, global
symmetries cannot be spontaneously broken in two
dimensions~\cite{Coleman:1973ci}, while the Higgs mechanism in two dimensions
itself is not prohibited.

The above construction of $\U(1)$ models can be generalized to four
dimensions. Our conclusion on the massive vector boson is similar, except the
point that now the models should be used with finite lattice spacings, because
the models are not renormalizable (in the first place, due to Yukawa couplings
with a compact Higgs field).

Finally, we comment on generalization to a non-abelian compact gauge group~$G$.
Natural generalization of the Higgs action is
\begin{equation}
   S_\kappa=\kappa\sum_x\sum_\mu\Real
   \tr\left\{1-\phi(x)^{-1}U(x,\mu)\phi(x+\hat\mu)\right\},
\end{equation}
where the compact Higgs field $\phi(x)$ is $G$-valued and the Higgs field
transforms as $\phi(x)\to\Lambda(x)\phi(x)$ under the lattice gauge
transformation. The fermion actions would be replaced by
\begin{widetext}
\begin{align}
   &S_{\text{light}}=\sum_x\left\{
   \overline\chi_+D_0\chi_++\overline\psi_-D\psi_-\right\},
\nonumber\\
   &S_{\text{mirror}}=\sum_x\left\{
   \overline\chi_-D_0\chi_-+\overline\psi_+D\psi_+\right\}
\nonumber\\
   &\qquad\qquad{}
   +y\sum_x\left\{
   \overline\chi_-R(\phi^{-1})\psi_+
   +\overline\psi_+R(\phi)\chi_-\right\}
   +h\sum_x\left\{
   \chi_-^TBR(\phi^{-1})\psi_+
   -\overline\psi_+BR(\phi)\overline\chi_-^T\right\},
\end{align}
\end{widetext}
where $B$ denotes the charge conjugation matrix. We assumed that the fermion
$\psi$ belongs to a unitary (generally reducible) representation~$R$ of $G$
and, $R(\phi)$, for example, denotes the Higgs field in that representation.
$\chi$ is a spectator (gauge singlet) and we have to introduce
$\dim R$~spectators. The lattice Dirac operators and the chirality projections
are defined according to the gauge representations of the fermions. We do
not write down an explicit form of gauge transformations, etc, because
generalization from the abelian case is obvious. Now, we may make change of
variables (that corresponds to the unitary gauge)
\begin{align}
   &U'(x,\mu)=\phi(x)^{-1}U(x,\mu)\phi(x+\hat\mu),
\nonumber\\
   &\psi'(x)=R(\phi(x)^{-1})\psi(x),\qquad
   \overline\psi'(x)=\overline\psi(x)R(\phi(x)).
\end{align}
Then the total action becomes independent of the Higgs field~$\phi$ and we have
the physical spectrum similar to that of the above $\U(1)$ case. Note that, in
this model, all vector bosons become massive and the $G$ gauge symmetry is
completely broken. The unitary gauge is equivalent to take $\phi(x)\equiv1$ and
this configuration is not invariant under any non-trivial gauge transformation.
Thus, with the above construction, it is impossible to leave some subgroup~$H$,
such as the $\U(1)_{\text{EM}}$ within the standard model
$\SU(3)\times\SU(2)_L\times\U(1)_Y$, unbroken.

The $G$-valued compact Higgs field precisely corresponds to the ``fundamental
representation'' case considered in Ref.~\cite{Fradkin:1978dv} and our
conclusion is consistent with that of~Ref.~\cite{Fradkin:1978dv}; the model is
in the Higgs phase. In two dimensions, because of the absence of a phase
transition in the pure gauge sector, an argument of Ref.~\cite{Fradkin:1978dv}
for the existence of the Coulomb phase does not apply. In four dimensions,
non-abelian models with massive vector bosons in which the mass is provided by
the St\"uckelberg (not Higgs in a limited sense) mechanism is not
renormalizable and the model should be used with finite lattice spacings.
On a related issue, see Ref.~\cite{Preskill:1990fr}.

In conclusion, the Yukawa-Higgs model with GW fermions proposed in
Ref.~\cite{Bhattacharya:2006dc} regrettably cannot be a starting point for
lattice formulation of chiral gauge theories, because the gauge symmetry is
always spontaneously broken.

\end{document}